\documentclass[prd,aps,onecolumn,11pt,floats,floatfix,nofootinbib,longbibliography,notitlepage,showkeys]{revtex4-2}

\usepackage{natbib}
\usepackage[T1]{fontenc}
\usepackage[utf8]{inputenc}
\usepackage{amsmath}
\usepackage{amsfonts}
\usepackage{amssymb}
\usepackage{multirow}
\usepackage{cases}
\usepackage[caption=false]{subfig}
\usepackage{graphicx}
\usepackage[hyperindex,colorlinks,allcolors=blue]{hyperref}

\newcommand{\bea}{\begin{eqnarray}}
\newcommand{\ena}{\end{eqnarray}}
\newcommand{\bean}{\begin{eqnarray*}}
\newcommand{\enan}{\end{eqnarray*}}

\begin{document}
\count\footins = 1000

\title{An overview of field theories of gravity}

\author{Mario Novello}\email{novello@cbpf.br}
\affiliation{Centro de Estudos Avançados de Cosmologia (CEAC/CBPF), Rua Dr. Xavier Sigaud 150, CEP 22290-180, Rio de Janeiro - RJ, Brazil.}

\author{Júnior D. Toniato}\email{junior.toniato@ufes.br}
\affiliation{Departamento de Química e Física \& Núcleo Cosmo-ufes, Universidade Federal do Espírito Santo (UFES), Alto Universitário s/nº, CEP 29500-000, Alegre - ES, Brazil.}

\date{\today}

\begin{abstract}
In the general relativity theory the basic ingredient to describe gravity is the geometry, which interacts with all forms of matter and energy, and as such, the metric could be interpreted as a true physical quantity. However the metric is not matter nor energy, but instead it is a new dynamical variable that Einstein introduced to describe gravity. In order to conciliate this approach to the more traditional ones, physicists have tried to describe the main ideas of GR in terms of standard conceptions of field theory. In this sense, curved metrics are seen as a dynamical variable emerging from a more fundamental field which lies upon a flat Minkowski spacetime. This was made by the hypothesis that the metric tensor may be written as $ g_{\mu\nu} = \eta_{\mu\nu} + h_{\mu\nu}$ where the tensor $ h_{\mu\nu}$ was interpreted either in terms of a spin-2 or constructed in terms of other fields. We review some proposals that were suggested in the treatment of gravity in terms of scalar, spinor and tensor fields configurations.
\end{abstract}


\maketitle

\tableofcontents

\section{Introduction}\label{sec:intro}
This review treat the different possibilities for describing gravitation in terms of a field theory and intends to show the relevance of these theories in the understanding of gravitational phenomena. Interestingly, only the scalar field had a first version prior to the theory of general relativity (GR).
Yet it was a rather simplistic version. Nevertheless, after the recognition of the importance of the modification of the metric of space-time made by Einstein several proposals to represent gravitational interaction in terms of modified geometry appeared using a scalar field to build an effective metric.

Although such a formulation of an effective metric has only developed in the last few decades, we can no doubt point to Gordon's proposal in 1923 as the original source of this idea \cite{Gordon}. Indeed, in the analysis of the propagation of electromagnetic waves in moving dielectrics, Gordon remarked that if one changes the geometry of the space-time in the domain of the propagation of a wave, the characterization of the motion can be described in a similar way as in the empty space, that is, the wave propagates through geodesics of this effective metric. At the first times this method of the effective geometry was used only for the propagation of waves. Later on, 
it was realized that it can be used formally to describe the dynamics of distinct formulations of non-linear field theories, as it will be discussed here.

This manuscript is organized as follows. In section \ref{sec:framework} we provide an overview of the origins and developments of the effective metric for different kinds of force, including gravity. Section \ref{sec:gsg} deals with the early proposals for scalar theories of gravity and with the recent developments of a scalar theory within an effective metric framework. In section \ref{sec:stg} a spinor theory of gravity is presented, as well as an unified treatment of gravity and Fermi (weak) interaction. Section \ref{sec:spin2} discuss the tensorial formulations, from the simplest reproduction of general relativity by Feynmann and collaborators to the most recent approaches. We conclude in Section \ref{sec:conclusion} with some comments on these alternative field theories of gravity.

\section{The main framework}\label{sec:framework}

Although general relativity is usually presented in the framework of Riemannian geometry, it is possible to fully describe the exact Einstein\rq s theory of gravity in terms of a spin-2 field propagating in an arbitrary background spacetime (see, for instance, Refs. \cite{GPP,Deser}). 
The main idea can be summarized as follows. Consider a flat Minkowski background (just to simplify our exposition) endowed with a metric $\eta_{\mu\nu}.$ In a Lorentzian coordinate system the metric of the background takes the standard constant expression. We may allow for general coordinates; the curvature tensor however vanishes,
\begin{equation}
	R^{\alpha}{}_{\beta\mu\nu}(\eta_{\varepsilon\sigma}) = 0.
\end{equation}
In a Galilean coordinate system the metric of the background can assume standard constant expression. From now on $\eta_{\mu\nu}$ are the component (not necessarily constant) of the Minkowski metric in general coordinates. Then one introduces a symmetric second order tensor $ h_{\mu\nu}$ and writes
\begin{equation}
	g_{\mu\nu} =  \eta_{\mu\nu} -  h_{\mu\nu}. \label{9julho12}
\end{equation}
This binomial form is an exact expression for the  metric $ g_{\mu\nu}.$ Note however that its inverse, the contravariant tensor $g^{\mu\nu}$, is not in general a binomial form but, instead, it is an infinite series,
\begin{equation}\label{inverse-inf}
	g^{\mu\nu} = \eta^{\mu\nu} +  h^{\mu\nu} - h^{\mu\alpha} \,h_{\alpha}{}^{\nu} + ...
\end{equation}

There are two main postulates founding general relativity: $i)$ The background Minkowski metric is not observable. Matter and energy interact gravitationally only through the combination $ \eta_{\mu\nu} - h_{\mu\nu}$ and its derivatives. Any test body in a gravitational field moves along a geodesic relative to the metric $g_{\mu\nu}$; $ii)$ The dynamics of gravity is described by an equation relating the contracted curvature tensor $ R_{\mu\nu} $ to the stress-energy tensor of matter.

In other words general relativity  assumes the hypothesis that gravity may be interpreted as nothing but a modification of the geometry of the underlying spacetime and the spacetime metric satisfies a dynamical equation that controls the action of matter on the modifications of the geometry. Although the first hypothesis have already demonstrated its success in describing a large number of phenomena, the second hypothesis has a more restrict support, it has a less sound acceptance. Indeed, it is well known the existence of a profuse number of models which modifies GR dynamics. Here we concentrate in those theories that describe the dynamics of the geometry as a consequence of other interactions due to non-gravitational process.

\subsection{Binomial metrics}

Recently, interest has grown in the study of a specific hypothesis that dynamical processes can be satisfactory described as a consequence of the modification in the geometric structure of spacetime. It is clear that, for GR specifically, there is a natural motivation once gravity is taken as an universal form of interaction. But in effective theories, the spacetime metric arises as a result of the dynamical equations of non-gravitational fields. The features of this procedure has been analyzed in the context of propagation of linear waves \cite{Hadamard} and to nonlinear theories \cite{novellobittencourt}.

The method is based on writing the spacetime metric in the form of Eq. \eqref{9julho12}
which gives, automatically, an infinite series for its contravariant version, as in Eq. \eqref{inverse-inf}. Thus, the usage of the same GR approach deals with serious difficulties. Alternatively, one can implement a novel interpretation and understand the theory as a spin-2 field propagating in a flat background. Meanwhile, one could also to consider the conditions under which the contravariant metric results in a binomial form as well.

Considering the metric structure, defined as
\begin{equation}
	g_{\mu\nu} = a\eta_{\mu\nu} + bh_{\mu\nu}.
\end{equation}
If $h_{\mu\nu}$ satisfies the closure relation below,
\begin{equation}
	\eta^{\alpha\beta} h_{\mu\alpha} \, h_{\beta\nu} = m \, \eta_{\mu\lambda} + n \, h_{\mu\lambda}.
	\label{closure}
\end{equation}
then the contravariant metric, $g^{\mu\nu}$, will also have a binomial form,
\begin{equation}
	g^{\mu\nu} = \alpha \,\eta^{\mu\nu} + \beta \, h^{\mu\nu},
\end{equation}
where $h^{\mu\nu}=\eta^{\mu\alpha}\eta^{\nu\beta}h_{\alpha\beta}$ and
\begin{align}
	\alpha = \frac{a + bn}{a(a + bn) - m b^2},\quad
	\beta = \frac{-b}{a(a + bn) - m b^2}.\label{beta}
\end{align}

\subsection{Newtonian attractions}

Let us face the first main proposal of general relativity, that is: is it possible to eliminate the acceleration induced on an arbitrary body $A$ by a gravitational field using a convenient modification of the metric associated to the spacetime where $A$ is propagating? We know the answer of general relativity: all gravitational effects are equivalently described in terms of the universal metric modification of the geometry of the spacetime.
Notwithstanding, let us show how the answer can be found in other approach. Note however that this must not be considered as an a priori once one could well make other kind of convention to ascribe specific metrics to different events.
If we accept this point of view then, each interaction, each event, each process requires a particular modification of the metric environment in such a way that one can eliminate any kind of force by just a modification of the metric in which a body moves in such a way that it is interpreted as a free body. In other words, the metric becomes just a convention to eliminate the force that drives the motion of the body.

The main direct consequence of this is that, contrary to the principles of general relativity, those metrics do not need any additional constraint that should be interpreted as its dynamics. The very fact that GR assumes that the universality of gravity is the origin to accept its geometric interpretation and, once this modification of the geometry is universal, completely independent of any particular process, it makes almost obligatory the existence of a particular dynamics for the metric. On the other hand, in the treatment of other forces, once each process has its own particular metric, in which the body movement is interpreted as inertial, there is no extra dynamics for the geometry: its characterization depends on the interaction.

A classical example is that of a photon propagating inside a moving dielectric, which may acquires an acceleration. In 1923 Gordon showed that it is possible to describe photon's path as a geodesics over a modified metric \cite{Gordon}.
This means that a change in the geometry can eliminate the acceleration felt by the photon. The interest on this description is that it allows its generalization for any kind of accelerated path, independently of the origin of the force and for any kind of massive or massless particle. Let us show the nature of this procedure by considering the simple case of an accelerated motion in flat Minkowski spacetime, when the acceleration is the gradient of a scalar function,
%
%
\begin{equation}\label{accel}
	a_{\mu} = \partial_{\mu} \Phi,
\end{equation}
with $\Phi$ an arbitrary scalar field.

Using the freedom in the definition of the four-vector we set $ \eta_{\mu\nu} \, v^{\mu} \, v^{\nu} = 1 $ and thus the acceleration is orthogonal to the velocity.
Let us construct for any congruence of curves the associated effective metric under the form
\begin{equation}
	\hat{q}^{\mu\nu}  =  \alpha \, \eta^{\mu\nu} + \beta \,  v^{\mu} \, v^{\nu},
	\label{153}
\end{equation}
with $\alpha$ and $\beta$ arbitrary (in principle) functions of $\Phi$.
The corresponding covariant expression is
\begin{equation}
	\hat{q}_{\mu\nu}  = \frac{1}{\alpha} \, \eta_{\mu\nu} - \frac{\beta}{\alpha \, (\alpha + \beta)} \,  v_{\mu} \, v_{\nu}.
	\label{154}
\end{equation}

We are using the hat symbol to denote objects in the auxiliary geometry where the metric is given by  $\hat{q}_{\mu\nu}.$ The associated covariant derivative of an arbitrary vector $ S^{\mu}$, represented by a semi comma $( ; )$, is defined by
\begin{equation}
	S^{\alpha}{}_{; \mu} = S^{\alpha}{}_{, \mu} + \hat{\Gamma}^{\alpha}{}_{\mu\nu} \, S^{\nu}
\end{equation}
where the corresponding Christoffel symbol is given by
\begin{equation}
	\hat{\Gamma}^{\varepsilon}{}_{\mu\nu} = \frac{1}{2} \, \hat{q}^{\mu\nu} \, \left( \hat{q}_{\lambda\mu , \nu} +\hat{q}_{\lambda\nu , \mu} -\hat{q}_{\mu\nu , \lambda}\right).
\end{equation}

By setting $\hat{v}^{\mu}= \Omega^{- 1} \, v^{\mu},$
and requiring that $ \hat{v}^{\mu}$ must be normalized, one obtains that
$\Omega =1/\sqrt{\alpha + \beta}.$
In order to identify this congruence generated by $ \hat{v}^{\mu}$ and $ v_{\mu}$  we require that $\Omega $  be constant along the motion, that is, $ v^{\mu} \, \partial_{\mu} \Omega = 0.$ The condition that makes the congruence as geodesic is provided by
\begin{equation}
	\partial_{\mu} \Phi = \hat{\Gamma}^{\varepsilon}_{\mu\nu} \, v_{\varepsilon} \, v^{\nu} = \frac{1}{2} \, ( \alpha + \beta) \, v^{\alpha} \, v^{\nu} \, \hat{q}_{\alpha\nu , \mu}.
\end{equation}
Then
\begin{equation}
	a_{\mu} + \frac{1}{2} \, \partial_{\mu} \ln (\alpha + \beta) = 0.
\end{equation}
Thus, for any congruence $\Gamma$ of accelerated curves in Minkowski spacetime driven by a potential as in \eqref{accel}, it is always possible to construct an associated auxiliary metric in the form \eqref{154} such that, if
\begin{equation}
	\alpha + \beta = e^{- 2 \Phi}.
\end{equation}
the paths of the curves become geodesics in this effective geometry.

More general situations of geometric modifications to describe accelerated motions can be found in Ref. \cite{novellobittencourt}. The method exposed before is in the basis of the so called analogue models of gravity. In these models, non-gravitational phenomena (the accelerated motion) are used to mimic typical kinematic effects that occurs in curved spacetimes (free motion of test particles in the presence of gravity). Most popular of these analogue models are the artificial black holes, where the effective metric presents an event horizon-like structure. For a review on the subject, the reader can refer, for instance, to Refs. \cite{novellovisservolovik,Cardoso:2013}.

In the  next sections we enter in a distinct analysis where different fields, obeying a given nonlinear equation of motion, can be  described in terms of modifications of the background geometry generated by themselves.

\subsection{Effective metrics in nonlinear scalar fields}\label{sec:scalarmetric}
Motivated by the discussion of the previous section, we wish now to discuss the whether is possible to translate nonlinear scalar field dynamics within a Minkowski spacetime into linear field equations over an effective geometry. Consider the following nonlinear Lagrangian,
\begin{equation}
	\label{lagr}
	L = V(\Phi) \,\omega,
\end{equation}
where $\omega\equiv\eta^{\mu\nu} \partial_{\mu}\Phi \, \partial_{\nu}
\Phi$, as before.
The field equation derived from the minimal action principle is given by,
\begin{equation}
	\frac{1}{\sqrt{-\eta}}\partial_{\mu} \, \left(\sqrt{-\eta}  \, \eta^{\mu\nu}\,
	\partial_{\nu} \Phi\right) + \frac{1}{2} \, \frac{V'}{V} \, \omega =0,
	\label{23julho3}
\end{equation}
where $ V' \equiv dV/d\Phi $ and  $ \eta $ is the determinant of $
\eta_{\mu\nu}.$

The field equation (\ref{23julho3}) can be seen as that
of a massless Klein-Gordon field propagating in a curved
spacetime whose geometry is governed by $\Phi$ itself.
In other words, the same dynamics can be written
either in a Minkowski background or
in another geometry constructed in
terms of the scalar field. Following the steps established in
\cite{Novello:2011sh}, let us introduce the contravariant metric tensor
$ q^{\mu\nu}$ by the binomial formula
\begin{equation}
	q^{\mu\nu} = \alpha \, \eta^{\mu\nu} + \frac{\beta}{\omega} \,
	\partial^{\mu}\Phi \,
	\partial^{\nu} \Phi,
	\label{9junho1}
\end{equation}
where $ \partial^{\mu} \Phi \equiv \eta^{\mu\nu}\,\partial_{\nu} \Phi$
and parameters $\alpha $ and $\beta$ are dimensionless
functions of $\Phi.$  Note that the quantity $w$ can be written in terms of its effective counterpart,
\begin{equation}
	\Omega  \equiv q^{\mu\nu}
	\partial_{\mu}\Phi \, \partial_{\nu} \Phi= (\alpha + \beta) \,\omega.
\end{equation}
From this expression, giving $ \alpha$ and $\beta $ we obtain $\Omega$ as function of $\omega$ and $\Phi$.

The corresponding covariant expression of the metric, defined as the
inverse $q_{\mu\nu} \, q^{\nu\lambda} = \delta^{\lambda}_{\mu}$, is
also a binomial expression, once it satisfies the closure relation \eqref{closure}. Its coefficients can be founded by inverting expressions in \eqref{beta},
\begin{equation}
	q_{\mu\nu} = \frac{1}{\alpha} \, \eta_{\mu\nu} -
	\frac{\beta}{\alpha \, (\alpha + \beta) \, \omega} \, \partial_{\mu} \Phi
	\, \partial_{\nu} \Phi.
	\label{9junho11}
\end{equation}

Now we ask whether it is possible to find $\alpha $ and $ \beta,$ in such a way
that the dynamics of the field\ (\ref{23julho3}) takes the form
\begin{equation}
	\Box \, \Phi = 0,
	\label{23julho5}
\end{equation}
where $\Box$ is the Laplace-Beltrami operator relative to the metric $q_{\mu\nu}$, that is
\begin{equation}
	\Box \, \Phi \equiv  \frac{1}{\sqrt{- q}}
	\partial_{\mu} ( \sqrt{- q} \,q^{\mu\nu} \,\partial_{\nu} \Phi).
\end{equation}
We then conclude that, given the Lagrangian $L = V(\Phi) \omega$ with an arbitrary potential $V(\Phi)$,
the field theory satisfying Eq. (\ref{23julho3}) in Minkowski spacetime is
equivalent to a massless Klein-Gordon field  in the metric
$q^{\mu\nu}$ provided that the functions $\alpha(\Phi)$ and $ \beta(\Phi) $ satisfy the condition
\begin{equation}
	\alpha + \beta = \alpha^{3} \, V.
	\label{9julho5}
\end{equation}
It is worth to note that such an equivalence is valid for any dynamics described in
the Minkowski background by a Lagrangian non linear in the kinetic term $\omega$  (see details in
\cite{Goulart:2011cb}).

Up to now, there is no relevant physical meaning associated with the effective metric $q_{\mu\nu}$. It determines a spacetime geometry from where the scalar field dynamics can be seen as a sourceless wave. The important step in the direction to attribute a physical significance to $q_{\mu\nu}$ is to determine what is its role when other fields are present. It can acquires an universal character, in the same lines as in GR, giving rise to a scalar field theory of gravity as it will be discussed in the next section.

\section{Scalar gravity}\label{sec:gsg}

\subsection{Preliminary attempts}
The history of the first attempts to describe gravitational phenomena with a single scalar field coincides with the history of the conception of general relativity by itself. Already in 1907, two years after the formulation of special relativity, Einstein had dedicate himself to develop a relativistic generalization of Newton's gravitational theory \cite{Einstein:1907}. His ideas can be summarized as follows. Firstly, there is the natural generalization of Newtonian force to a four-vector quantity,
\begin{equation}\label{qf}
	{\cal F}^\mu = \frac{d}{d\tau}\left(m\,\frac{dx^\mu}{d\tau}\right)\,,
\end{equation}
where $m$ is the particle mass and $\tau$ is the proper time. This last definition imposes an auxiliary condition due to the constancy of light velocity.
Contracting \eqref{qf} with the metric, and assuming that $M$ is always a constant, one obtains the relation
\begin{equation}\label{cond}
	\eta_{\mu\nu}{\cal F}^\mu\frac{dx^\nu}{d\tau}=0\,.
\end{equation}
This condition is naturally satisfied in electromagnetism, once the electromagnetic four-force is defined as ${\cal F}^\mu_{em}=F^{\mu\alpha}\eta_{\alpha\beta}{dx^\beta}/{d\tau}$, and due to the anti-symmetry of Maxwell's tensor, $F^{\mu\nu}$.

For the gravitational case, the simplest way to follow is to construct the gravitational four-force from the gradient of the scalar potential $\Phi$, and generalize Poisson's equation to a four dimensional one. Thus, one come up with the following equations,
\begin{eqnarray}\label{fg}
	{\cal F}^\mu_g = mc^2\,\partial^{\,\mu}\Phi\,,\quad
	\square\,\Phi = -\frac{4\pi G}{c^2}\,\rho\,.
\end{eqnarray}
In the above, $\square$ is the d'Alembert operator and $\rho$ is the mass density distribution. Note that the scalar potential is now dimensionless. However, with the condition \eqref{cond}, one gets ${d\Phi}/{d\tau}=0\,.$
Such result basically implies that the gravitational force on any particle is always null, an inconsistency pointed out by Einstein and which lead him to question the compatibility between gravity and special relativity. The discussion that emerged in the following years, concerning mainly the equivalence principle and general covariance, ended up with the formulation of the general theory of relativity. The details of this enticing history can be seen, for instance, in Ref. \cite{Norton:1992}. But in the mean time, a feel other attempts to develop a scalar theory of gravity was made by Nordstr\"{o}m.

\subsubsection{Nordstr\"{o}m's theories}\label{sec:nordstrom}
In order to improve Einstein's developments on the relativistic generalization of Newtonian gravity, Nordstr\"{o}m explored the idea of a variable mass \cite{Nordstrom:1912,Nordstrom:1913}. Equation (\ref{fg}) is then rewritten as follows,
\begin{equation}\label{eq2}
	m\,\frac{dv_\mu}{d\tau}+v_\mu\,\frac{dm}{d\tau}=\, mc^2\,\frac{\partial\Phi}{\partial x^\mu}.
\end{equation}
Contracting the above equation with $v^\mu$ and integrating gives the gravitational field dependence of the mass,
\begin{equation}\label{massa}
	m(\Phi) =\, m_{0}\,e^{\Phi},
\end{equation}
with $m_{0}$ a constant. Folowing that equation (\ref{eq2}) return the equation of motion of test particles, namely
\begin{equation}\label{eq3}
	c^2\,\frac{\partial\Phi}{\partial x^\mu}=\, \frac{dv_\mu}{d\tau}+v_\mu\,\frac{d\Phi}{d\tau}\,.
\end{equation}

Although the independence of equation \eqref{eq3} with respect to the mass could suggest an agreement of this theory with the weak version of the equivalence principle, a simple case of study can reveal some inconsistencies. Consider a free fall in a static gravitational field acting only in the direction of the movement, say $\Phi=\Phi(z)$. From \eqref{eq3} one can obtain the relations
\begin{equation}
	\frac{dV_z}{dt}= -c^2\left(1-\frac{V^2}{c^2}\right)\frac{d\Phi}{dz}\,,\quad \frac{dV_x}{dt}=0\,, \quad \frac{dV_y}{dt}=0\,,
\end{equation}
with $V_i$ are the components of the tree-vector $\vec{V}=d\vec{x}/dt$ and $t$ is the local time. Thus, bodies with horizontal velocities would affect the vertical free fall, which shows that Nordstr\"{o}m model violates the equivalence principle.

Another point of criticism in the Nordstr\"{o}m theory concerns the source of the gravitational field. The mass density $\rho$ is a projection of the energy momentum tensor dependent on the observer,
\begin{equation}
	\rho=T_{\mu\nu}v^\mu v^\nu,\quad\mbox{with}\quad 	T_{\mu\nu}=-\frac{2}{\sqrt{-\eta}}\frac{\delta (\sqrt{-\eta}\,L_m)}{\delta\eta^{\mu\nu}},
\end{equation}
and $L_m$ is the Lagrangian function of matter fields. One thus see that the modified Poisson equation in \eqref{fg} is not Lorentz invariant.
Actually, it was Einstein that suggested that the only scalar quantity that could be source of the gravitational field is the trace of the energy momentum tensor, $T=T_{\mu\nu}\eta^{\mu\nu}$. In a second model, Nordstr\"{o}m implemented this modification together with the argument that the theory should be nonlinear, once the energy momentum tensor should also account for the gravitational energy. As a result, the field equation and the gravitational force law are now described as follows,
\begin{equation}\label{n1}
	\square\,\Phi=-\frac{4\pi G}{c^4}\,\frac{T}{\Phi}\,,\qquad F^\mu_g=\frac{mc^2}{\Phi}\,\partial^{\,\mu}\Phi\,.
\end{equation}
The equation of motion for test particles is also modified, namely
\begin{equation}\label{n2}
	c^2\,\frac{\partial\Phi}{\partial x^\mu}=\, \Phi\,\frac{dv_\mu}{d\tau}+v_\mu\,\frac{d\Phi}{d\tau}\,,
\end{equation}
with the mass now having a linear dependence with the gravitational field, i.e. $m(\Phi) =\, m_{\scriptscriptstyle 0}\,\Phi$.

In 1913, Einstein described Nordstr\"{o}m's second theory,  together with his own \textit{Entwurf} formulation, as the only ones with a satisfactory description of the gravitational interaction \cite{einstein2}. However, the \textit{Entwurf} model were not a scalar theory of gravity, being the first time that Riemannian manifolds and tensor calculus were used to explore the idea of curved spacetimes as a consequence of gravity \cite{entwurf}.\footnote{\textit{Entwurf} means outline, or sketch, and it is how the 1913 paper by Einstein and Grossmann is usually known \cite{entwurf}. It is a short name derived from the original title \textit{Entwurf einer verallgemeinerten relativit\"{a}tstheorie und einer theorie der gravitation}.} Both theories would satisfies basics requirements such as the equality of inertial and gravitational mass, reduction to special relativity as a limiting case and conservation of energy and momentum.

Nowadays, it is clear that Nordstr\"{o}m's theory lacks of empirical confirmation once it cannot provide any deflection of light and it does not explain the anomalous motion of Mercury. However, at that time, there wasn't any measurement of light bending and no other theory could satisfactory explain Mercury's perihelion advance.

\subsubsection{Einstein and Fokker reformulation of Nordstr\"{o}m theory}\label{sec:einstein-fokker}

Nordstr\"{o}m second theory also have kinematic effects that are typical in metric theories of gravity: the gravitational field would alter the tick of clocks and the length of rods. Such feature indicates that Nordstr\"{o}m scalar theory could be interpreted as spacetime theory. It was Einstein and Fokker that demonstrated this property applying the mathematical tools of tensor calculus to Nordstr\"{o}m theory \cite{einstein3}. It is shown that equation of motion (\ref{n2}) describes a geodesic in a curved geometry conformal to Minkowski spacetime,
\begin{equation}\label{metric}
	g_{\mu\nu}= \Phi^2\,\eta_{\mu\nu}\,.
\end{equation}
In this case, the Ricci scalar is simply described as $R= -({6}/{\Phi^3})\square\,\Phi$. On the other hand, the energy-momentum tensor should now be defined as a variation of matter Lagrangian with respect to the physical metric $g_{\mu\nu}$,
\begin{equation}
	\label{tmunu_g}
	T_{\mu\nu}=-\frac{2}{\sqrt{-g}}\frac{\delta (\sqrt{-g}\,L_m)}{\delta g^{\mu\nu}}.
\end{equation}
Thus, the trace appearing in Nordstr\"{o}m field equation [cf. \eqref{n1}] will transform as $T \rightarrow\, {T}/{\Phi^4}$. Therefore, theory's dynamics can be rewritten as follows,
\begin{equation}\label{escalar}
	R(g)= \,\frac{24\pi G}{c^4}\,T(g)\,.
\end{equation}
where the notation $R(g)$ and $T(g)$ is used to emphasize that both the Ricci scalar and the trace of energy-momentum tensor are constructed with the physical metric $g_{\mu\nu}$.

This formulations describes the gravitational field in a covariant and geometrical way, which guarantees the validity of the equivalence principle in its strong and weak versions. It is worth to mention that within this description of Nordstr\"{o}m theory, the mass are no longer dependent of the gravitational field. However, spacetime would admit preferred coordinate systems in which
equation (\ref{escalar}) reduces to Nordstr\"{o}m equation. Moreover, in metric (\ref{metric}), only the conformal factor has a dynamics influenced by matter and energy, but its structure cannot be changed by simply modifying the matter distribution of a system.

Even so, Einstein's work on Nordstr\"{o}m theory was essential for him to return to the quest of finding a fully covariant theory of gravity. The outcome of his quest, the well know theory of general relativity, led to the replacement of the traditional single scalar potential of gravity in favor of the ten independent components of the spacetime metric. The history of this remarkable endeavor is depicted, for instance, in Ref. \cite{janssen}.

\subsubsection{Scalar theories post general relativity}
The observational and experimental success of general relativity cemented the concept that gravity is a geometrical phenomenon. Any theoretical formulation aiming to viably describe gravitational interactions should be a metric theory of gravity. In this sense, the majority of scalars theories of gravity proposed after the advent of general relativity were also metric theories of gravity. In general, those can be grouped in two classes: conformally flat theories and stratified theories.

The conformally flat class of scalar theories are formulations where the physical metric ($g_{\nu\mu}$) is generated from a scalar field ($\Phi$) and flat metrical structure ($\eta_{\mu\nu}$) through a conformal relation,
$g_{\mu\nu}=\Omega(\Phi)\eta_{\mu\nu}.$
Different forms of the function $\Omega(\Phi)$ and the field equation of $\Phi$ will give rise to distinct theories. Nordstr\"{o}m theory is an example of a conformally flat theory with $\Omega = \Phi^2$.

Stratified theories considers, besides a scalar function representing the gravitational field, a universal time coordinate $t$ whose gradient is covariantly constant and timelike with respect to the flat background, i.e.
$t_{;\mu;\nu}=0$ and $\eta^{\mu\nu}t_{,\mu}t_{,\nu}=-1.$
The function $t$ defines a preferred reference frame in a which the spatial slices of spacetime are conformally flat. In a frame where $t_{,\mu}=\delta_\mu^0$, the physical metric is constructed according to the line element below,
\begin{equation}
	ds^2=g_{\mu\nu}dx^\mu dx^\nu= f(\Phi)dt^2 - h(\Phi)\eta_{ij}dx^idx^j,
\end{equation}
with $f(\Phi)$ and $h(\Phi)$ arbitrary functions that will differ one theory from another.

Conformally flat theories cannot produce an coupling between gravity and electromagnetism since Maxwell's equations are conformally invariant. This implies that there is no light bending effect in those theories. The preferred frame idea brought within stratified theories is a way to avoid this complication. Other possibility explored to circumvent this problem is to consider light velocity dependent of gravitational field $\Phi$ (see, for instance, Ref. \cite{page}. Notwithstanding, the effective metric formalism discussed in section \ref{sec:framework} lead the way to another possibility for describing gravity through a single scalar field. The fundamentals of this scalar field theory of gravity is discussed in the next section.

\subsection{Geometric scalar gravity}
The geometric scalar gravity (GSG) was first presented in Ref. \cite{novellobittencourtmoschellagoulartsalimtoniato} and it is based upon the idea of giving an universal and gravitational meaning for the effective scalar metric discussed in previous section [cf. \eqref{154}]. In section \ref{sec:framework}, it was discussed two founding postulates of GR that determine all gravitational phenomena to be interpreted as a modification of spacetime geometry given by Einstein's equation. Although those postulates can be adjusted, the main ideas brought by them are fundamental for what is called metric theories of gravity. We start this section reviewing the postulates of Nordstr\"{o}m-like theories of gravity, and then proceeding to the postulates of the geometric scalar gravity.

\subsubsection{Postulates of Nordstr\"{o}m-like theories}
Nordstr\"{o}m's second theory and its geometric reformulation by Einstein and Fokker was discussed in sections \ref{sec:nordstrom} and \ref{sec:einstein-fokker}. We can recast any Nordstr\"{o}m-like theory as based on the following postulates:
\begin{enumerate}
	\item Gravity is mediated by a single massless scalar field $\Phi$, satisfying a linear dynamics in Minkowski spacetime.
	
	\item All kind of matter and energy interact with $\Phi$ only through a minimal coupling with the ``physical" metric $g_{\mu\nu}$, which is conformal to the Minkowski metric,$g_{\mu\nu} = A^2(\Phi) \eta_{\mu \nu}\,.$
\end{enumerate}

The hypothesis that the matter fields are minimally coupled to the physical metric $g_{\mu\nu}$ warrants the validity of the weak equivalence principle.
The total action leads to the following field equations:
\begin{eqnarray}
	\Box \Phi =  -\frac{4\pi G}{c^4} A' A^3\, T, \label{fieldeq} \quad\mbox{and}\quad
	T^{\mu\nu}_{\ \ \ ;\nu} = 0\,,\label{2}
\end{eqnarray}
with $A'=dA/d\Phi$\,. Here the energy-momentum tensor of the non-gravitational fields, $T_{\mu\nu}$, is defined as in equation \eqref{tmunu_g}, with its trace given by $T_{\mu\nu}g^{\mu\nu}$. The covariant derivative in Eq. (\ref{2}) is taken with respect to $g_{\mu\nu}$. The scalar curvature of the conformal geometry reads $	R = -\, {6\, \Box A}/{A^3}.$
By using this relation, equation (\ref{fieldeq}) may be rewritten in the Einstein-Fokker covariant form,
\begin{eqnarray}
	R =\ \frac{24\pi G}{c^4}\, A'^2\, T - \frac{6 \,A''}{A} \,g^{\mu\nu}\partial_\mu \Phi\partial_\nu\Phi \,. \label{fieldeq2}
\end{eqnarray}
For $A=\Phi$, equation \eqref{fieldeq2} recovers the original one from Nordstr\"{o}m's second theory in the Einstein-Fokker reformulation [cf. \eqref{n1}].

\subsubsection{Postulates of geometric scalar gravity}
As a possible way to solve the difficulties that exist in Nordstr\"om's theories due to a priori assumptions of conformal symmetry, the GSG theory  explore the possibility that the physical metric be related to the Minkowski metric by a more general structure as in \eqref{9junho1}. This hypothesis immediately permits the coupling of the electromagnetic field  to the physical metric. It is also explored more general Lagrangians for the gravitational part of the action and, in particular, modifications of the kinetic term.

Summarizing,  the GSG  enlarge the Nordstr\"om's family according to the following postulates:
\begin{enumerate}
	\item Gravity is mediated by a scalar field $\Phi$ satisfying a nonlinear dynamics in Minkowski's spacetime, described by the action,
	\begin{equation}
		S_{\text {gravity}} = \frac{1}{\kappa c} \int L(\Phi, \partial _\mu\Phi) \,  \sqrt{ -\eta}  \,d^4x  \,. \label{disf2}
	\end{equation}
	
	\item All kind of matter and energy interact with $\Phi$ only through a coupling with the physical metric $q_{\mu\nu}$, given by
	\begin{equation}
		q_{\mu\nu} = a(\Phi,\omega) \eta_{\mu \nu} + b(\Phi,\omega) \partial_\mu \Phi \partial_\nu \Phi \label{disformal}
	\end{equation}
	where $\omega =\eta^{\mu\nu} \partial_\mu \Phi \partial_\nu \Phi$. Thus, the action for the matter may be written as
	\begin{eqnarray}
		S_{\text{matter}}(\psi,q_{\mu\nu})=\frac{1}{c}\int \sqrt{-q}\,L_m\,d^4x.
		\label{disformalgsg}
	\end{eqnarray}
	with $q$ the determinant of $q_{\mu\nu}$.
\end{enumerate}

When $a= a^{2}(\phi)$, $b = 0$ and $L$ is the Lagrangian of a massless Klein-Gordon field we are back to Nordstr\"om theories. It is worth to note that the coupling of matter and energy fields with a metric structure like in \eqref{disformal} have been considered in many contexts within scalar-tensor theories of gravity (see, for instance, Refs. \cite{beke,Sakstein:2014isa,Brax2018a}). They are usually called disformal gravity. Moreover, general ways to deform the spacetime has been considered to enlighten many gravitational problems and metric disformations are also embedded in this class of transformations (see \cite{Capozziello2008b} and the references therein).
However, the GSG model seems to be the unique one where a scalar field disformally coupled is used to modify the original Nordstr\"om's idea and describe gravity in the context of a purely scalar theory.

We call a theory belonging to this family a geometric scalar theory of gravity for the obvious reason that matter interacts with gravity only through coupling to the physical metric (\ref{disformal}). A general geometric scalar theory of gravity is characterized by three functions: the functions $a$ and $b$ characterizing the metric and the Lagrangian $L$ of the scalar field.

\subsection{A particular case of GSG}
It is clear that functions $a,b$ and $L$ cannot be chosen arbitrarily, once the resulting theory must be in agreement with gravitational tests. In this section, it is discussed some basic assumptions used as a guide to propose a model within GSG.

We first note that, when $T=0$,  Nordstr\"om's  theories  coincide with the flat space massless Klein-Gordon theory irrespectively of the conformal factor (see Eq. (\ref{fieldeq})). A similar -- but also distinct -- feature is shared by a particular class of geometric scalar theories of gravity that in vacuum reduce to the massless Klein-Gordon equation but now w.r.t. the curved spacetime physical metric $q_{\mu\nu}$. This possibility was considered in section \ref{sec:scalarmetric}, thus, by restricting the metric structure to the particular form of \eqref{9junho1}, and with a Lagrangian $L=V(\Phi)\omega$, the vacuum dynamics of the scalar field take the form of a massless Klein-Gordon field, $\Box\Phi=0$, provided that condition \eqref{9julho5} holds.
Therefore, contrary to the case of Nordstr\"{o}m's theory, in the GSG approach the field equation keeps its nonlinearity and the gravitational scalar field is self-interacting.

Consider now the Newtonian correspondence principle. For simplicity, we will set units such that $G=c=1$. Assuming the hypothesis, as in general relativity, that test particles follow geodesics relative to the metric $q_{\mu\nu}$, in the case of a static weak field configuration and low velocity motions, we have
\begin{equation}
	\frac{d^{2} x^{i}}{dt^{2}} \approx - \, \Gamma^{i}_{00}\approx - \, \frac{1}{2} \partial^{i} (\ln
	\alpha)= - \,
	\partial^{i} \, \Phi_{N},
	\label{22julho1}
\end{equation}
where $\Phi_N$ is the Newtonian gravitational potential. Thus, the relation between the time component of the physical metric and the Newtonian potential is given by
\begin{equation}
	q_{00} = \frac{1}{\alpha} \approx 1 + 2 \,\Phi_{N}.
\end{equation}
Moreover, using that $\Box\Phi=0$, one obtains the right (vacuum)
Newtonian limit, $\nabla^2\Phi_N=0$. The above relation shows the linear behavior of the metric coefficient $\alpha$ in order to be in agreement with Newtonian limit. However, for analytical purposes, we will explore the
consequences of extrapolating this relation and take into account a more
general expression, namely
\begin{equation}
	\label{alpha_phi}
	\alpha = e^{- 2 \,\Phi}.
\end{equation}

To determine the dynamics of $\Phi$ in a presence of matter fields, one first vary $S_{\rm matter}$ with respect to the physical metric, introducing then the energy-momentum tensor. After some algebra (see Ref. \cite{novellobittencourtmoschellagoulartsalimtoniato} for details), the final form of the theory's field equation reads
\begin{equation}\label{eqgsg}
	\sqrt{{V}} \, \square\Phi= \kappa \, \chi\,,
\end{equation}
where the right hand side is simplified by the notation
\begin{equation}
	\chi=\,-\,\frac{1}{2}\left[\,T +\left(2 -\frac{V'}{2V}\right)E +C^\lambda_{~\,;\lambda}\right]\,.
\end{equation}
\begin{equation}
	E \equiv \frac{T^{\mu\nu} \, \partial_{\mu}\Phi \, \partial_{\nu}\Phi}{\Omega}\,,
\end{equation}
\begin{equation}
	C^{\lambda}\equiv\frac{(\alpha^2\,V-1)}{\Omega} \, \left( T^{\lambda\mu} - E \, q^{\lambda\mu} \right) \, \partial_{\mu}\Phi\,.
\end{equation}
The symbol ``$\,;\,$'' stands for the covariant derivative with respect to the physical metric $q_{\mu\nu}$. The above equation makes clear that not only the trace of the energy-momentum tensor acts as a source for the gravitational field, but also  a non-trivial coupling between the gradient of the scalar
field  and the complete energy-momentum tensor
of the matter field . This makes possible the coupling between gravity and  electromagnetism in this scalar theory.

The coupling constant $\kappa$ is determined through the correspondence principle with Newton's gravitational theory. The Newtonian limit can be obtained when $T^{00}\approx \rho$, $T^{0i}\approx 0$ and $\partial_t\Phi\approx0$, where $i$ stands for spatial components and $\rho$ is the matter density. Applying this approximation scheme gives the identification $\kappa=8\pi G/c^4$.

The next task is to determine the functional dependence of $\beta$
on $\Phi$,  or either the form of the potential $V(\Phi)$ once both functions are related by \eqref{9julho5}.
To select among all possible Lagrangians of the above form we look for indications from the various circumstances in which reliable test have been performed. In this vein, one must analyze the consequences of GSG for the solar system and, in a first approximation, the requirement that vacuum and spherically solution meets Schwarzschild geometry could be a strong indication in that direction. This requirement is achieved by the following potential,
\begin{equation}
	V(\Phi) = \frac{(\alpha-3)^2}{4 \, \alpha^3} \label{alp_sch}.
\end{equation}
Details of how to obtain this result can be found in Ref. \cite{novellobittencourtmoschellagoulartsalimtoniato}. It has been shown
that three charts are necessary to cover the Schwarzschild solution and that the region inside the horizon is somehow surprisingly the disformal transformation of a Euclidean metric \cite{novellobittencourtmoschellatoniato-2}.

\subsection{Post-Newtonian analysis}
The post-Newtonian analysis of a gravitational theory consists of a practical way to obtain theoretical predictions that can be confronted with observations. Essentially, it is a weak-field and slow motion approximation scheme that, at first order, it guarantees that the theory recovers Newton's gravity results and, on higher orders, introduces the departures from the Newtonian formulation. After solving the approximated field equations, the metric components are presented as a combination of Newtonian potential and several others post-Newtonian ones. Within this method, there is the parametrized post-Newtonian (PPN) formalism,  a useful tool to test metric theories of gravity since it gives a physical meaning to each one of the metric coefficients appearing in the  post-Newtonian expansion, i.e. the PPN parameters. Also, it extracts observational bounds for nine of the ten parameters used in the formalism (see Ref. \cite{will-book} for a detailed presentation of PPN).

The GSG model considered in this review is not covered by the PPN formalism due to the non-standard potentials that appears in the expanded metric.\footnote{The limitations of the PPN formalism is rather recurrent for modern alternative theories of gravity (see, for instance, the discussion in Refs. \cite{toniato1,toniato2}).} Even so, in Ref. \cite{novellobittencourtmoschellatoniato-2} it was investigated the post-Newtonian limit of this particular GSG model and it is verified that, in a static monopole approximation, the expanded metric does fit with the general metric of the PPN formalism. Within that simplification, the possible PPN parameters that can be read off are the following ones,
\begin{align}
	\beta=1, \ \ \gamma=1,\ \ \xi=0,\ \
	\zeta_2=0,\ \  \zeta_3=0,\ \  \zeta_4=-\frac{4}{3}\,.
\end{align}

The physical meaning of these parameters in the standard PPN formalism are the following. The $\gamma$ is associated with light trajectories and it is bounded by light deflection and Shapiro time-delay effects. The latter  is the one which brings the strongest bound on that parameter, $\vert\gamma-1\vert \lesssim 10^{-5}$. With $\gamma$ constrained, $\beta$ happens to be the single parameter relevant for the perihelion shift of Mercury, and the consequent bound is $\vert\beta-1\vert \lesssim 10^{-5}$. The Whitehead parameter $\xi$ is related with preferred-location effects and its bound comes from spin-precession measurements of millisecond pulsars, which gives $\xi\lesssim 10^{-9}$. The $\zeta$'s parameters, when different from zero, are manifestations of non conservation of total energy and momentum. The $\zeta_2$ and $\zeta_3$ are constrained by binary system and lunar acceleration effects, with $\zeta_2\lesssim10^{-5}$ and $\zeta_3\lesssim10^{-8}$. The $\zeta_4$, related with matter pressure effects, is not a independent parameter due to theoretical constraints that gravity produced by kinetic energy, internal energy and pressure should satisfy, and it does not have an independent bound.

However, it is important to note that there is no guarantee that these parameters keep their original physical meaning within GSG formulation, and the direct usage of the PPN bounds is not adequate here (or in any other theory which does not fit the PPN formalism). Notwithstanding, the fact that the values of $\beta,\gamma,\xi,\zeta_2$ and $\zeta_3$ are the same as the ones obtained in general relativity could represent a strong indication of a satisfactory description of post-Newtonian physics. A not restrictive study of the post-Newtonian equations of motion is necessary in order to confirm such expectations.  The not vanishing of  $\zeta_4$ may not be viewed as inconsistent since the proper definition of gravitational energy in GSG is ambiguous. For instance, due to the non trivial relation between the scalar field source term $\chi$, and a generic energy-momentum tensor, there is a class of symmetric tensors $t^{\mu\nu}$ which satisfies $\chi(t^{\mu\nu})=0$. Those tensors may have influence on the energy and momentum conservation statements while not impacting scalar field dynamics.

\subsection{Gravitational waves}
The new era of multi-messenger astronomy, with direct observations of gravitational waves (GW), brings significant improvement for testing gravitational theories. In particular, the velocity of GW in vacuum has been measured to be the same of light \cite{Abbott_2017}, which is consistent with the theoretical prediction from GSG, as it was shown in Ref. \cite{Toniato:2019}. A weak field approximation shows that the perturbed scalar field satisfies
a wave equation over Minkowski's spacetime. Such solutions yields an oscillatory behavior for the metric as GW propagating in spacetime.

Concerning the polarization states, it is well know that any metric theory of gravity can have at most six distinct polarization modes. These modes are related to the so called Newmann-Penrose quantities (NPQ), deduced from the irreducible parts of the Riemann tensor \cite{Newman:1961qr}. However, the theory's dynamics can reduce the number of modes by vanishing some of the NPQ. As an example, two transversal polarization modes of general relativity is due to the fact that only one of the NPQ is not identically null. Sometimes, the polarization modes are observer-dependent (which is not the case of GR), as described by the classification presented in Ref. \cite{Eardley:1974nw}, and this happens to be the case of GSG. Once the Ricci scalar is not identically null in vacuum, one gets the most general class of the $E(2)$-classification with an always present longitudinal mode of polarization and the detection of the others five states being observer-dependent. This is a substantial distinction from general relativity.

The existence of any mode of polarization can not be extract from the recent detections of GW \cite{LIGOScientific:2019fpa,Abbott:2018lct}. Thus, GSG cannot be constrained in such aspect. In the future, with the increasing of network of detectors with different alignment, polarization states can be used to restrict gravitational theories. It is worth to note that $f(R)$ theories is also classified as GSG \cite{Alves:2009eg}.

However, GSG can be constrained by observational data from pulsars through its prediction for the orbital variation of a binary system that should be caused by the loss of energy due to gravitational radiation. The discussion on how to define an energy-momentum tensor for the linear GW, following a field theoretical point of view, starts by identifying the second order vacuum field equation for the approximation scheme $\Phi \approx \phi_{\scriptscriptstyle 0} + \phi_{\scriptscriptstyle (1)} + \phi_{\scriptscriptstyle (2)}$, yielding
\begin{equation}\label{fe2}
	\Box_{\eta}\phi_{\scriptscriptstyle (2)}= -\,\frac{9-\alpha_{0}}{3-\alpha_{0}}\,w_{\scriptscriptstyle (2)}.
\end{equation}
The sub-indexes indicates the order of smallness and $w_{\scriptscriptstyle (2)}=\eta^{\mu\nu}\partial_{\mu}\phi_{\scriptscriptstyle (1)}\partial_{\nu}\phi_{\scriptscriptstyle (1)}$ and $\Box_{\eta}$ is the d'Alembertian operator constructed with Minkowski metric.
The right hand side of this equation contains only the derivatives of the first order field $\phi_{\scriptscriptstyle (1)}$, thus it can be interpreted as the source for the second order field generated by the linear waves. However, an ambiguity emerges since GSG fundamental equation includes a non trivial interaction between matter/energy and the scalar field. In other words, the above equation must be recast in similar form as in \eqref{eqgsg}, leading to a non unique expression for the linear gravitational energy-momentum tensor. Specifically, any tensor $\Theta_{\mu\nu}$ with its second order approximation given by
\begin{equation}\label{tet}
	\Theta_{{\scriptscriptstyle (2)} \mu\nu}= \frac{1}{\kappa}\left(\sigma w_{\scriptscriptstyle (2)}\eta_{\mu\nu} +\lambda\,\partial_\mu\phi_{\scriptscriptstyle (1)}\,\partial_\nu\phi_{\scriptscriptstyle (1)}\right),
\end{equation}
with the constants $\sigma$ and $\lambda$ satisfying the relation,
\begin{equation}\label{c1}
	\sigma\left(\frac{9-5\alpha_{0}}{3-\alpha_{0}}\right) - \lambda\left(\frac{2\alpha_{0}}{3-\alpha_{0}}\right)=\frac{9-\alpha_{0}}{\alpha_{0}^{3/2}}\,,
\end{equation}
will give origin to a source term
\begin{equation}\label{f3}
	\chi_{\scriptscriptstyle (2)}(\Theta_{\mu\nu})= - \dfrac{9-\alpha_{0}}{2\kappa \alpha_{0}^{3/2}} \,w_{\scriptscriptstyle (2)}\,, \quad \mbox{and} \quad \sqrt{V_0}\,\Box_{\eta}\phi_{\scriptscriptstyle (2)}= \kappa\chi_{\scriptscriptstyle (2)}(\Theta_{\mu\nu}).
\end{equation}
Thus, $\Theta_{{\scriptscriptstyle (2)} \mu\nu}$ can be interpreted as the energy-momentum tensor for the linear gravitational waves in GSG. In the above expression, $\alpha_{0}$ is the metric coefficient calculated with the background field $\phi_0$ using definition \eqref{alpha_phi}.

The ambiguity in the GW energy-momentum tensor is encoded in the constant parameter $\lambda$, and has directly consequences for a system energy-loss rate due to the emission of gravitational waves. Specifically, for a binary system, the consequent average variation in the orbital period is determined by the following expression
\begin{align}\label{per}
	\frac{\dot{T}}{T}= \ \frac{3\lambda}{4}\frac{G^3 M^{3}}{m_1m_2\,a^{4} c^{5}}\,\ e^{2}(1-e^{2})^{-7/2}\bigg[f^{2} + \bigg(\dfrac{f^{2}}{4} +4Mf + 4M^{2}\bigg)e^{2} +2M^{2}e^{4}\bigg],
\end{align}
where $m_1$ and $m_2$ are the masses of each component of the system, $a$ is the semi-major axis, $e$ is the eccentricity, $M=m_1 +m_2$ and $f=5m_{1}m_{2}/{M}- 4M$ (details in Ref. \cite{Toniato:2019}). The expression above for the orbital period variation depends on $G$, $a$ and $c$ the same way as occur in GR, but the mass and eccentricity dependence are rather more involved.

It is possible to use data from the PSR 1913+16 binary pulsar to constraint the theory's free parameter $\lambda$.  However, although the orbital parameters of the binary system are extracted in a theory-independent way, the mass values estimation of the two bodies in the binary system are model-dependent \cite{Damour_2009}. Once that GSG has a satisfactory agreement with GR concerning Solar System tests at the first post-Newtonian order, one can use (as a first approximation) the mass values obtained in GR. Notwithstanding, deviations from the GR mass values will lead to a distinct numerical value for $\lambda$, but it would not necessarily invalidate how GSG describes the orbital period variation of binary systems.
The results are $\lambda= -1.111 \pm 0.003$ and $\sigma=3.444\pm0.002$ [after using condition \eqref{c1}], which completely fixes the energy momentum tensor expression for the gravitational waves in GSG. The minus sign in $\lambda$ is due to the fact that the orbital period is decreasing while the system emits gravitational waves, i.e., $\dot T<0$.

\subsection{Cosmological scenarios}
Assuming the scalar field $\Phi$ depends only on time, it yields a homogeneous and isotropic geometry,
\begin{equation}
	ds^{2} = dt^{2} - a(t)^{2} \, (dx^{2} + dy^{2} + dz^{2}).
	\label{him}
\end{equation}
with the cosmological scale factor being $a = e^\Phi$. Because the gravitational field depends only on $t$, it is natural to expect that all the relevant quantities have only temporal dependence too.
Assuming the material content of the universe can be modeled as a perfect fluid decomposed in terms of the cosmic observers, $v^{\mu}=\delta^{\mu}_0$ , the resulting cosmological equation,
\begin{equation}\label{jr}
	{a}|3a^2-1|\left(\frac{\ddot a}{a}+2\frac{\dot a^2}{a^2} \right) = \kappa\left[3p +\left(\frac{2\rho}{3a^2-1}\right)\right],
\end{equation}
where $\rho$ is the matter density ($c=1$) and $p$ is the fluid pressure.
There are therefore two regimes classified by the sign of the quantity $3a^2-1$, however, the time evolution respects that sign and GSG cosmologies, based on a perfect fluid, are divided into two classes. We call the solutions belonging the first class (i.e. solutions such that $3a^2>1$) Big Universes (BU), and the solutions belonging to the the second one, Small Universes (SU).  The adjective ``small" makes allusion to the fact the scale factor takes values in a compact interval, but also when this occurs the spatial section is of course infinite. The existence of two classes of solution is  a consequence of the choice of the potential (\ref{alp_sch}).

For barotropic fluids, when $p=\lambda\rho$, one has $\rho=\rho_0\, a^{-3(1+\lambda)}$ with $\rho_0$ constant. With a simple integration, both first and second time derivative of the scale factor can be written as a function of $a$, namely
\begin{equation}\label{vel}
	\dot a^2=\frac{M}{a^4} - 2\kappa\rho_0\, \frac{a^{-2-3\lambda}}{|3\,a^2- 1|}\,,
\end{equation}
where $M$ is a constant of integration, and
\begin{equation}\label{ace}
	\ddot a=-2\,\frac{M}{a^5} +\,\frac{\kappa\rho_0}{a^{3(1+\lambda)}}\left[\frac{2+3\lambda}{|3\,a^2 -1|} \pm\,\frac{6\,a^2}{(3\,a^2 -1)^2}\right]\,,
\end{equation}
with the upper sign for BU and the lower sign being valid for SU solutions.
As it should be expected, the above expression is singular at  $3a^2 =1$.
However this value is unattainable because within BU (SU) $\dot{a}^2$ becomes zero at a minimal (maximum) value $a_m$ strictly grater (smaller) than $1/\sqrt{3}$. Therefore big universes and small universes are two disjoint classes of cosmological solutions of GSG.
\begin{figure}[t]
	\centering
	\includegraphics[width=.32\textwidth]{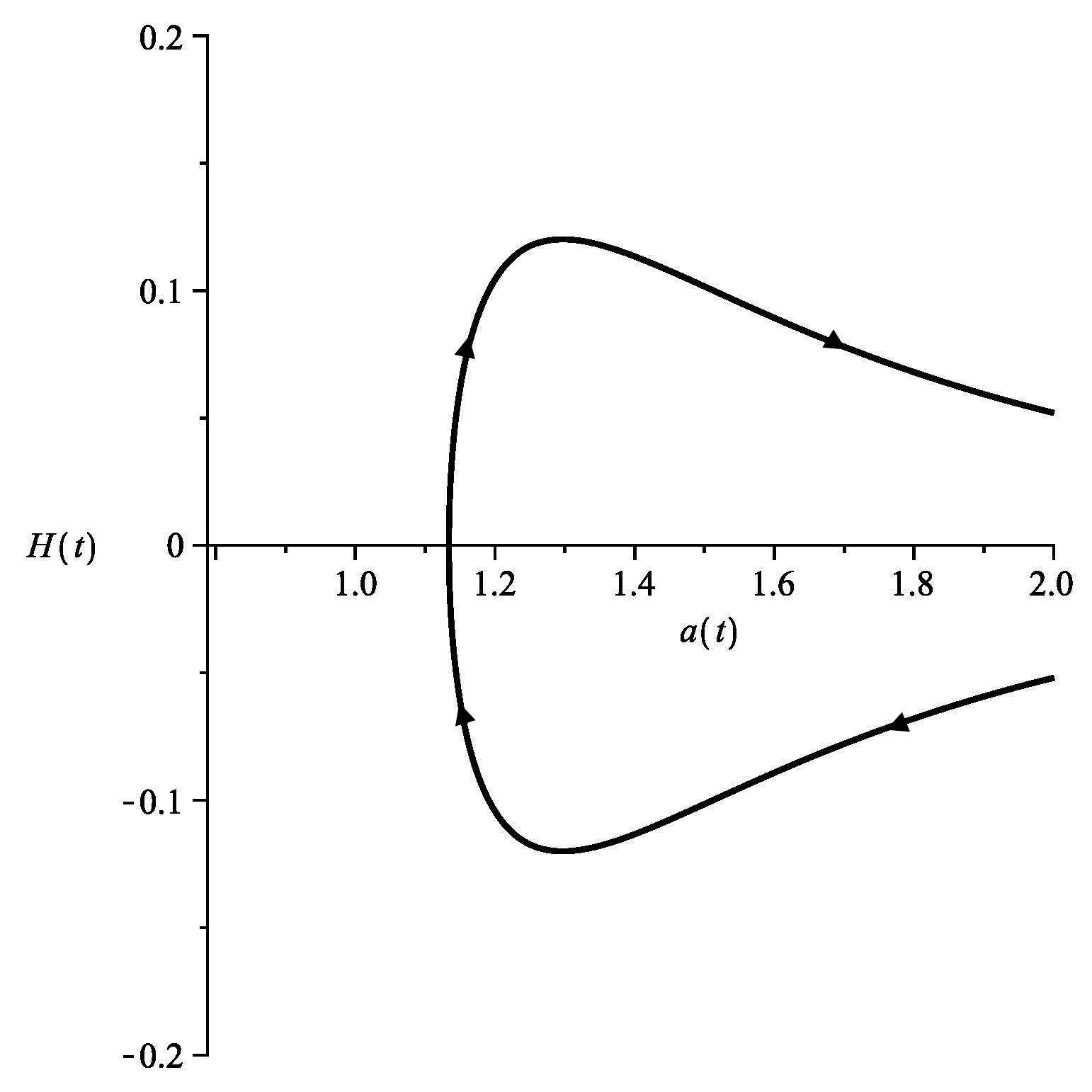}
	\hfill
	\includegraphics[width=.32\textwidth]{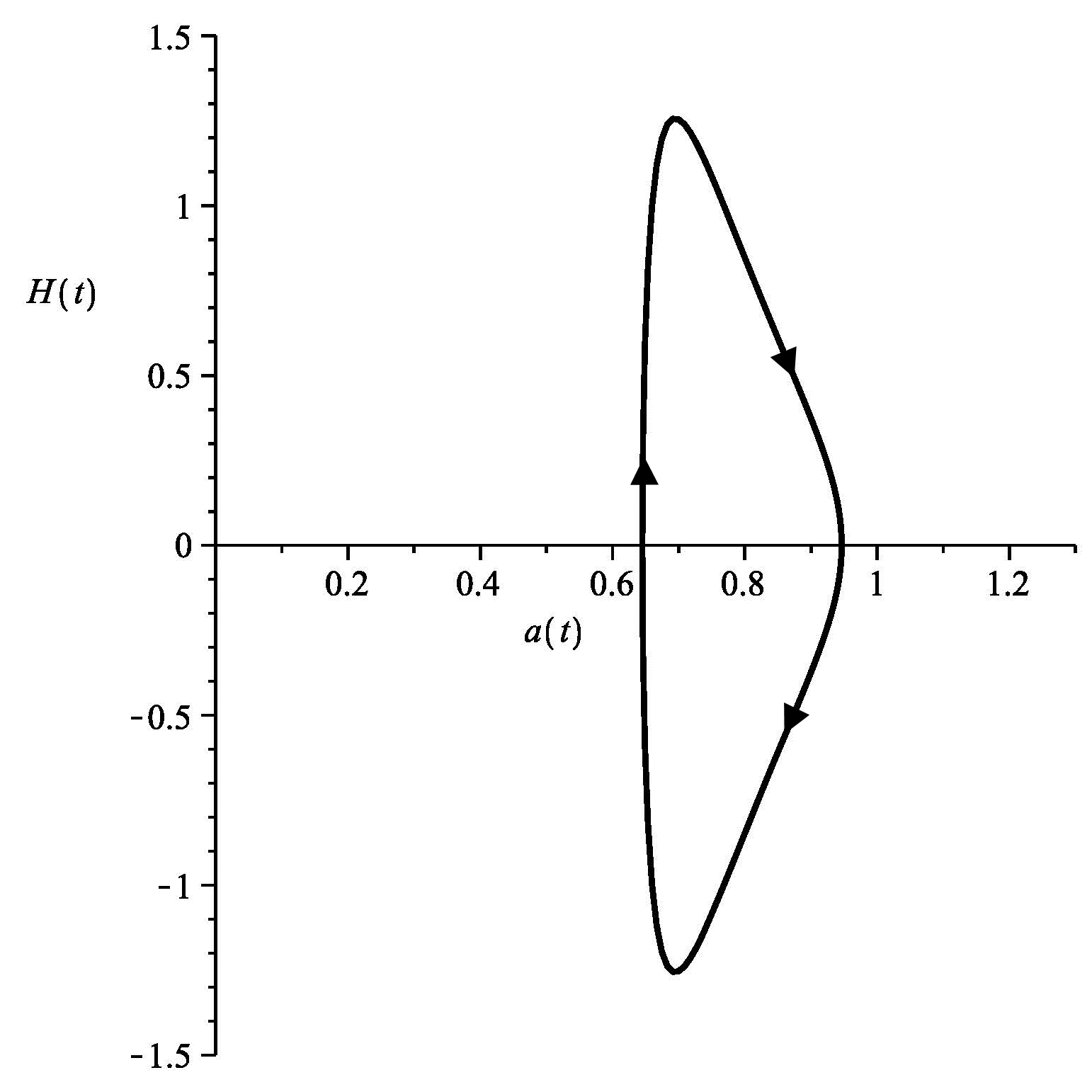}
	\hfill
	\includegraphics[width=.32\textwidth]{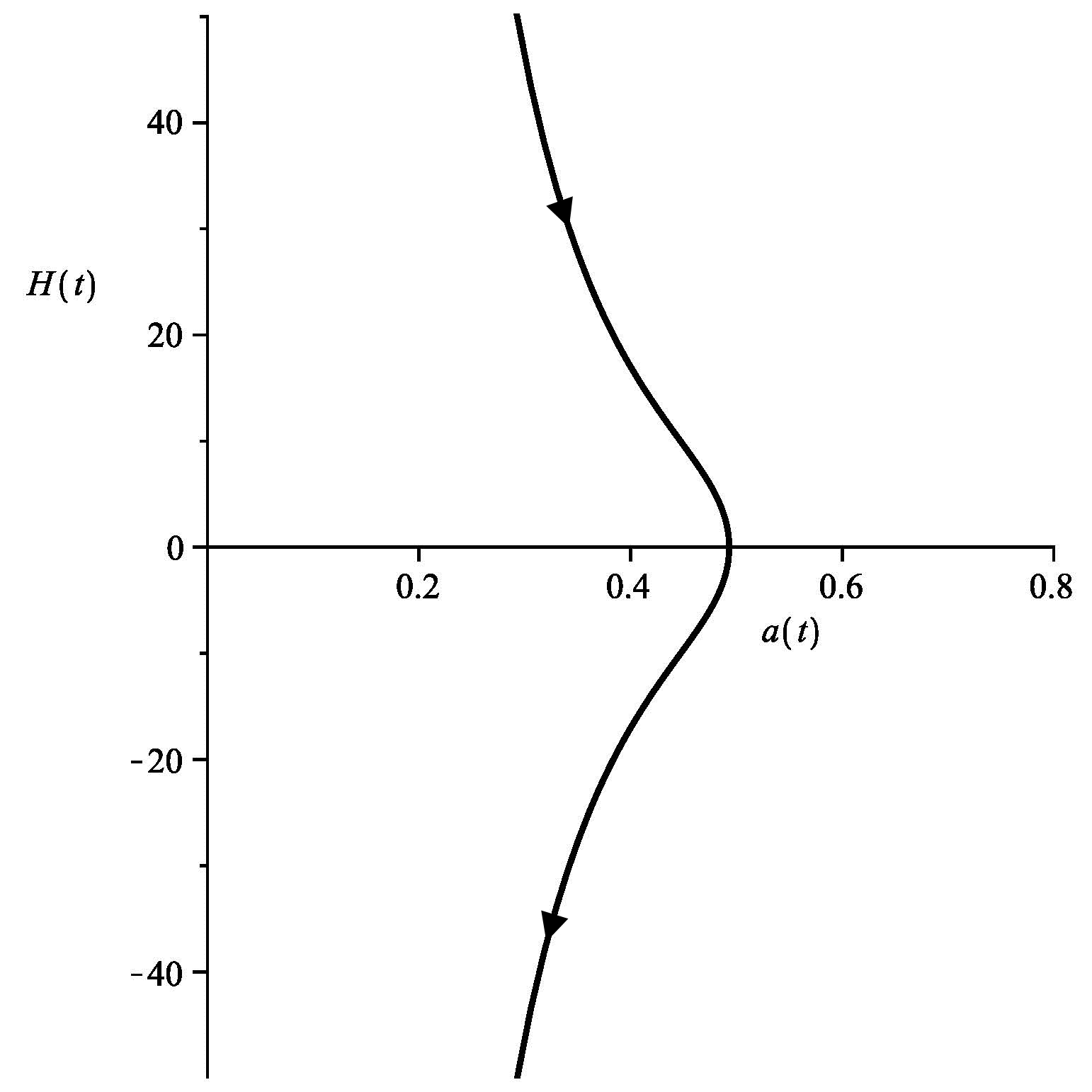}\\
	\hspace{2cm} (a) \hfill (b) \hfill (c) \hspace{1.5cm}
	\caption{Phase diagrams for  $M=0.9\,\kappa\rho_0\,$. (a) Big Universe filled with dust ($\lambda=0$). The universe is non singular and this property is shared with any fluid with positive $\lambda$, including radiation. (b) Big Universe cyclic scenario for negative values of $\lambda$. The figure shows the case of $\lambda=-1$ but the behavior is similar for any $\lambda<0$. (c) Dust filled Small Universe. There is a singularity in $a=0$ from where the evolution starts and ends after reach a maximum scale.}
	\label{fig:df}
\end{figure}

A qualitative analysis can be performed using equations \eqref{vel} and \eqref{ace} by plotting phase diagrams as shown in Figure \ref{fig:df}. The left image shows the BU solution for pressureless matter $(\lambda=0)$. This universe filled with dust does not have any singular behavior and presents an accelerated phase near de bounce. At late times, the universe decelerates. Fluids with positive $\lambda$ (including radiation) have a similar behavior. They all have a bounce, followed by an early accelerated phase and a final decelerated phase. The expansion last forever.

For negative values of $\lambda$ it is possible to have a static solution in BU for a particular choice of $M$. For the others possible values of $M$ the universe is cyclic as, for instance, the case of $\lambda=-1$, shown in Figure \ref{fig:df}(b).

The SU branch includes a class of solutions, for $\lambda<2/3$, which presents a singularity at $a=0$. In these scenarios there is an initial big bang followed from a decelerated phase. The universe reaches its maximum scale and then stops entering in contracting phase that unavoidably ends in the singularity at $a=0$. This case is shown in Figure \ref{fig:df}(c) for $\lambda=0$. For $\lambda>2/3$ in SU is possible to have a static solution for a specific value of $M$ or a cyclic and non-singular universe.

The possibility of having a bouncing for standard fluids (in BU) is quite remarkable. In Freedmann-Lema\^{\i}tre-Robertson-Walker (FLRW) cosmology the bouncing is possible either by non-minimal coupling with matter fields or by negative pressures. Indeed, in order to have an extremum of the scale factor $a(t)$ the expansion factor $\theta= 3  \dot{a}/a$ must vanish and its derivative be positive.
As it is well known, in GR the Raychaudhuri equation implies that this is possible when $(\rho + 3 p)/2<0$. Here the situation is different: the universe always bounces for pressure values equal or greater to zero. On the other hand, GSG cosmology does not have natural mechanisms to generate a late phase of accelerated expansion of the universe if only working with ordinary matter and energy contents.

More details of background cosmological solutions in GSG can be found in Ref. \cite{novellobittencourtmoschellatoniato}. A discussion of scalar perturbations, focusing on the growth of density perturbations in a matter dominated later decelerated phase was also performed in that reference. It was verified that gravitational instabilities are possible in GSG. Fortunately, they are slight different from GR, which means that the GSG cosmology can be tested separately with a properly analysis of data in this new framework.

\section{Spinor theory of gravity}\label{sec:stg}

For some deep thinkers of the past certain relations between numbers of physical phenomena may suggest a more intimate connection between them. For instance the observation of the Large Number Hypothesis of Dirac, . In the same vein, Stuckelberg argued that relationship of gravitational force and weak (Fermi) processes led him to propose a more deep connection between gravity and Fermi interactions.It is clear that these ideas do not prove the existence of such connections, but it permit the emergence of new ideas on physics as, for instance, in the case of Dirac, the development of the scalar-tensor theory of gravity.

The proposal to develop a theory of gravity that is based on the dynamics of spinor fields was inspired in the possible connection between gravity and the weak (Fermi) interaction based on a remark that can be made concerning the numerical values of the fundamental constants characterizing both interactions. In fact, the idea to extract motivation from such a ``nomerological'' argument dates back to Dirac's Large Number Hypothesis, where properties of the universe and elementary particles led him to propose the time dependence of the gravitational constant \cite{Dirac:1937}. A similar analysis is performed in Refs. \cite{Onofrio:2013,novellohartmann} in the context of the weak and gravitational interactions, suggesting that the geometrical interpretation of the gravitational phenomena can be seen as emerging from more fundamental spinor fields. In the next sections we will review recent developments in this direction.

\subsection{Fermi interaction and Gravity: the fundamental spinors}

In the case of Fermi interaction, the weak current is defined as
$C_{\mu} = J_{\mu} - I_{\mu},$
where $J_\mu=\overline{\Psi}_g\gamma_\mu\Psi_g$ and $I_\mu=\overline{\Psi}_g\gamma_\mu\gamma_5\Psi_g$.
Thus the Lagrangian for the interaction of two fermions, $ \Psi_{e} $ and $ \Psi_{n}$, can be written as
$ L_{Fermi} = g_{w} \, C_{\mu}^{e} \, C^{\mu}_{n} $.
The use of a parity violation term (the vector minus the axial vector currents in the Fermi interaction) guides the form of fundamental vectors that are treated as the basis of the gravitational metric. Indeed, the hypothesis made in the spinor theory of gravity (STG) is based on the existence of two fundamental massless spinorial fields which are the true elementary structure of gravity. These two fields  $ \Psi_{g}$ and $\Omega_{g}$ generate an effective metric which is the one dealing in general relativity. In other words, the proposal of GR that gravity deals with metric\rq s modification of space-time has a substratum that is identified with these gravitational fermions (g-fermions).

The main ideas of the STG can be synthesized in the following sentences:
\begin{enumerate}
	\item{The geometry of general relativity is an effective representation of two physical spin-1/2 fields, called \textit{g-fermions} and denoted by $ \Psi_{g}$ and $\Omega_{g}$, living in the Minkowski spacetime;}
	\item{$\Psi_{g}$ and $\Omega_{g}$ interact via Fermi process;}
	\item{Both fields couple with all forms of matter in an universal way;}
	\item{ This interaction can be described as a modification of the background Minkowski metric into an effective one $ g_{\mu\nu},$ following the fundamental ideas of general relativity;}
	\item{The effective metric $ g_{\mu\nu}$ does not have a dynamics by its own but inherits the dynamics of the fundamental fields $ \Psi_{g}$ and $\Omega_{g}.$ }
\end{enumerate}

The spinorial fields $ \Psi_{g}$ and $\Omega_{g}$ they both interact universally with all kind of matter and energy only through the metric $g_{\mu\nu}$, thus following the main principle of GR.
The gravitational metric $ g_{\mu\nu}$ is written in terms of null vectors that can be constructed with $\Psi_{g}$ and $\Omega_{g}$.
A combination of these vectors provide a dimensionless term $ h_{\mu\nu}$, 
\begin{align}
	h_{\mu\nu} = \ell\left[\Delta_{\mu} \, \Delta_{\nu} + \Pi_{\mu} \, \Pi_{\nu}+ \varepsilon \left( \Delta_{\mu} \, \Pi_{\nu} + \Delta_{\nu} \, \Pi_{\mu}\right)\right],
	\label{27abril1}
\end{align}
where $\ell$ is a constant with dimensions of length squared and
\begin{align}
	\Delta_{\mu}=& \frac{1}{\sigma \sqrt{J}}(J_\mu-I_\mu),\quad \mbox{with} \quad J^2 =J_\mu J^\mu,\\
	\Pi_\mu=& \frac{1}{\sigma \sqrt{j}}(j_\mu-i_\mu), \quad \mbox{with} \quad j^2=j_\mu j^\mu.
\end{align}
In the above expressions $j_\mu=\overline{\Omega}_g\gamma_\mu\Omega_g$ and $i_\mu=\overline{\Omega}_g\gamma_\mu\gamma_5\Omega_g$.
The $\sigma$ is a constant with dimensions of length, expected to be dependent of fundamental constants present in the weak interaction formulation, while $\ell$ should have influence of gravitational constant as well. Moreover, in order to satisfy the closure relation \eqref{closure}, one must have $\varepsilon^{2} = 1$.

It has been explored in the literature the simpler case  in which $\Omega_{g}$ vanishes, leading to
\begin{equation}
	h_{\mu\nu} = \ell\, \Delta_{\mu} \, \Delta_{\nu},
	\label{7maio21}
\end{equation}
and a simple inspection shows that such tensor satisfies the property
\begin{equation}
	h_{\mu\nu} \, \eta^{\nu\alpha} = h_{\mu\nu} \, g^{\nu\alpha},\quad \mbox{and}\quad g^{\mu\nu} = \eta^{\mu\nu} + h^{\mu\nu}.
	\label{25abril1}
\end{equation}
By dimensional analysis, the constant $\sigma$ can be defined as $ \sigma^2 = \kappa \hbar c$, where $ \kappa = 8 \, \pi G/c^4$. Thus, fixing $\ell=\kappa \hbar c$, $h_{\mu\nu}$ becomes dimensionless. It is worth to note that the $\Delta_\mu$ could be simply defined to be dimensionless, but the method worked above makes usage of the connection between weak and gravitational interactions, starting with a vector depending on constants from the weak interaction realm and still getting a dimensionless tensor by combining with another quantity constructed with constants typical of the gravitational interaction.

\subsection{The gravitational field}

The fundamental spinors live in the flat Minkowski background where the metric $ \gamma_{\mu\nu}$ is defined from the fundamental objects,
\begin{equation}
	\gamma_{\mu\nu}= \frac{1}{2} \, (\gamma_{\mu} \, \gamma_{\nu} + \gamma_{\nu} \, \gamma_{\mu}),
\end{equation}
which is a multiple of the identity of the Clifford algebra. In the case $ \Omega_{g} = 0$ the dynamics of $ \Psi_{g}$ is given by the Dirac equation in arbitrary coordinate systems,
\begin{equation}
	i\gamma^{\mu}
	\nabla_{\mu} \, \Psi_g = 0
	\label{26maio1}
\end{equation}
and the internal connection contains beyond the conventional Fock-Ivanenko term an additional one,
\begin{equation}
	\Gamma_{\mu} = \Gamma_{\mu}^{FI} + U_{\mu}\,.
	\label{5abril5}
\end{equation}

In STG the covariant derivative of the $\gamma^{\mu}$ is given by the Riemannian condition
\begin{equation}
	\gamma_{\mu ; \nu} = [U_{\nu}, \gamma_{\mu} ]
	\label{5abril2}
\end{equation}
where $ U_{\mu} $  besides to be a vector is an arbitrary object from the associated Clifford algebra. There is a simpler case where
\begin{equation}
	U_{\mu} = \frac{1}{4} \, \gamma_{\mu} \, \gamma^{\alpha} \, U_{, \, \alpha}
	\label{5abril3}
\end{equation}
with $ U_{,\alpha} = \partial_{\alpha} U$. The vector $U_\mu$ is driven by an external scalar field $H$ such that
$U =  \varepsilon \, H\,.$
This form implies the Riemannian condition for the metric, that is,
$\gamma_{\mu\nu ; \lambda} = 0$ and the factor $ 1/4 $ is just for latter convenience. For the origin of this term see Ref. \cite{novellohartmann}. Note that in the case we use Euclidean coordinates, the gamma\rq s are constant and the Christoffel symbol vanishes, as well as the Fock-Ivanenko connection.
Moreover, due to the usage of only one spinor field, as well as the fact that $ \Delta_{\mu}$ is a null vector, both the determinant of $g_{\mu\nu}$ and $\eta_{\mu\nu}$ coincide. Important consequence arises in such limited framework, describing, for example, a spherically symmetric and static spacetime and a cosmological scenario (see Ref. \cite{novellohartmann}).


\subsection{The effect of gravity on matter}
Following the idea brought by general relativity, gravitational interaction is described by the substitution of the flat background by a more general structure governed by $\Psi_{g}$. To illustrate this procedure, consider the case in which matter is represented by a massless scalar field, $\Phi$. The matter action will be
\begin{equation}
	S_m = \frac{1}{2}\int \, \sqrt{-g} \, g^{\mu\nu} \, \partial_{\mu} \Phi \, \partial_{\nu} \Phi\,.
	\label{16abril3}
\end{equation}
Once the metric $g_{\mu\nu}$ depends on the spinor field, the variation of the total action, with respect to $\Psi_g$ and $\Phi$, will give, respectively
\begin{equation}
	i \, \gamma^{\mu} \, \nabla_{\mu} \Psi_g + \kappa\, E_{\mu\nu} \, Q^{\mu\nu}\,\Psi_g = 0\quad\mbox{and}\quad\Box \Phi = 0,
	\label{27maio5}
\end{equation}
with
\begin{equation}
	E_{\mu\nu} = \Phi_{, \, \mu} \, \Phi_{, \, \nu} - \frac{1}{2}  \Phi_{, \,\alpha} \, \Phi_{, \, \beta} \, g^{\alpha\beta} \, g_{\mu\nu},
\end{equation}
being the energy-momentum tensor of the scalar field and
\begin{equation}
	Q^{\mu\nu} =  \left(\frac{g_{w}}{J^{2}}\right)^{1/4} \,  \Delta^{\mu} \,  \gamma^{\nu}\,(1 - \gamma_{5}) \,  - \frac{1}{2J^{2}}\,  \Delta^{\mu} \, \Delta^{\nu}  \, (A + i B \gamma_{5}), \label{Q}
\end{equation}
with $A=\overline{\Psi}_{g}\Psi_{g}$ and $B=i\overline{\Psi}_{g}\gamma_5\Psi_g$.
It is worth to note that the non-linear term, $ A + i B \gamma_{5}$ , present in the dynamics of $\Psi_g$ is not a particularity of this model. Such a term is also present in dynamical models and field theories of elementary particles (see, for instance, Refs.  \cite{heisenberg,heisenberg2,nambu}).

The generalization to other forms matter is directly obtained through the usual variation of a matter action with respect to the metric $g_{\mu\nu}$, yielding
\begin{equation}
	i \, \gamma^{\mu} \, \nabla_{\mu} \Psi_g \,+\, \kappa \, T_{\mu\nu} \, Q^{\mu\nu} \, \Psi_g = 0\,,
	\label{27maio4}
\end{equation}
Given the expression of the energy-momentum tensor $T_{\mu\nu}$, this equation yields the value of $\Psi_g$ that allows the construction of the current $\Delta_{\mu}$ and then the metric $g_{\mu\nu}$. In this sense,  it plays the same role as the equation that relates the Ricci curvature to the energy-momentum tensor in general relativity.

\subsection{Local and global vacuum solutions}
We now review some results obtained in STG for two of the simplest scenarios for gravitational theories: the geometry surround a spherically symmetric and static mass distribution; an isotropic vacuum cosmology.

For the spherically symmetric vacuum case, it is found the following line element
\begin{equation}
	ds^2 = \left( 1 - \frac{1}{S} \right) \,dt^2 - \left(1 - \frac{1}{S}\right)^{-1} \, dr^2 - r^2 \, d\theta^2 - r^{2} \, \sin^2\theta \, d\varphi^2,
	\label{20abril 6}
\end{equation}
where $S = r \, ( a_{0} - 2 \, \lambda \, \log r).\label{S}$
In the limit when there is no self-interaction term ($\lambda=0$), the line element above reduces to the usual Schwarzschild form, as can be verified through \eqref{S}. This suggest one to set $ a_{0} = 1/ r_H,$ with $r_H=2 m$. It is worth to note also that the limiting case where $r_H$ goes to zero, $1/S$ vanishes and Minkowski spacetime is recovered. In conclusion, the vacuum solution of the non-interacting spinor theory reproduces the traditional Schwarzschild spacetime.

The new physics emerges when the spinors does self interact. There is an extra singular behavior of the metric when $r$ is equal to $r_s=e^{a_0/2\lambda}$. Once the Ricci scalar at this radius is $R=8\lambda/S^3$, one can see that it is a true singularity. This fact makes this particular non-linear solution unfeasible in general. Even so, one could use it, in certain cases, to obtain some insight of how the spinor self-interaction does modify the gravitational interaction. It could then be considered a satisfactory description of a vacuum region when the source have a radius greater than $r_s$. This situation can be more easily obtained in the limiting case $\lambda<0$ and $\vert\lambda\vert\ll a_0$, where $r_s\ll 1$. If one goes further and also assumes a weak field regime, i.e. $r\gg m$,  the line element can be approximated to
\begin{equation}
	ds^2\approx\left( 1 - \frac{2 m}{r} - \frac{8\lambda m^2}{r}\ln r\right)dt^2 - \left( 1 + \frac{2 m}{r}  +\frac{8\lambda m^2}{r}\ln r\right) \, dr^2 -r^2d\Omega^2.
\end{equation}

The horizon, $R_H$, in this metric occurs when $S(r)=1$, as can be seen from \eqref{20abril 6}. The consequent equation has solution for $\lambda\neq 0$ and it is given in terms of a Lambert $W$ function.\footnote{The Lambert $W$ function is also called omega function or product logarithm.} Within the limit of negative and small $\lambda$, one has
\begin{equation}
	R_H\approx 2m\Big[1- 4\vert\lambda\vert m\ln(2\vert\lambda\vert)\Big].
\end{equation}
For completeness of this discussion, we note that if $\lambda$ is positive the position of the singularity goes to higher values of the radius, i.e. $r_s\gg 1$. In the case where $\lambda\gg a_0$, one has $r_s\approx 1$ either for positive or negative values of $\lambda$. Also, the line element will be no longer a small deviation from Schwarzschild, since $S(r)\approx 2\lambda r\ln r$.

Other new result concerns the existence of an isotropic vacuum  solution of the fundamental equations of STG, which reads
\begin{equation}
	ds^2 = dt^2 - \tanh^2 t \, \left(dX^2 + dY^2 + dZ^2 \right)\,.
	\label{22maio2}
\end{equation}
This result of STG should be compared with the case of GR where in the absence of matter is not possible to find an isotropic cosmology. Moreover, one can note that, as $t\rightarrow \infty$ the solution is regular and it tends to Minkowski spacetime, in contrast with the singular behavior of Kasner metric.

\section{Tensor theory}\label{sec:spin2}

Although general relativity is presented as an universal modification of the metrical
properties of spacetime, an alternative way to describe GR as a spin-2 field theory in the
same lines as any other interaction was revived by some authors (see, for instance, Refs. \cite{Deser,GPP}). The idea goes back to the works of Gupta \cite{gupta}, Kraichnan \cite{Krai55} and others investigation
on spin-2 field. In his book, {\it Lectures on Gravitation}, Feynman exposed this idea in a very simple way \cite{feynman}. It was shown that a field theoretical approach of gravity should be
possible and its basic ingredients should deal, besides the spin-2 field
$\varphi_{\mu\nu}$ with two metric tensors: an auxiliary one $\gamma_{\mu\nu}$ --- which is
not observable --- and an effective one, $g_{\mu\nu}$, related by
$g_{\mu\nu} = \gamma_{\mu\nu} + \varphi_{\mu\nu}$.

The basic hypothesis of GR concerns the extension of the
equivalence principle beyond its original domain of experimental evidence,
that concerns material substance of any form, the adoption of its validity not only by matter or
non-gravitational energy of any sort but also by gravity energy itself.
Such an universality of interaction is precisely the cornerstone that
makes possible the identification of a unique overall geometry of spacetime
$g_{\mu\nu}$. The properties of gravity are associated
to the Riemannian curvature, which becomes then the equivalent substitute of
gravitational forces. What we can learn, from the approach of Feynman et al, is that such geometric scheme is permissible but it is by no way mandatory, that is, the geometrical description of GR
is nothing but a choice of representation. All observable characteristics
and properties of Einstein theory can be well described in terms of a spin-2 field $\varphi_{\mu\nu}$. We emphasize that such an alternative description of GR in no way sets a
restriction on it, but only enlarges its power of understanding.

From this approach it follows that contrary to a widespread belief,   the field-theoretical way of
treating GR appeals to a two metric structure. Furthermore, Feynman has shown that the coherence of a
spin-2 theory that starts with the linear Fierz-Pauli equation,
written in terms of the symmetric field $\varphi_{\mu\nu}$, in a Minkowskian
spacetime requires, in a very natural way, due to the self-interaction process
described above, the use of an induced metric tensor, the quantity
$g_{\mu\nu}$ \cite{Fierz}. This is the standard procedure. Nevertheless,
and just by tradition, this is not the way that Einstein theory is presented in text books.\footnote{The absence in the literature of such alternative but equivalent way to present Einstein	theory of gravity seems to be the main responsible for the young students of	theoretical physics to understand GR as a completely separate and different theory from any other field. In textbook presentations of GR one makes the choice of	a unique geometry. This, of course, does not preclude an alternative equivalent description \cite{GPP}.} This interpretation allows us to state that two-metric theories of gravity are less exotic than it is usually displayed \cite{will}. Let us emphasize that the second metric is nothing but a convenient auxiliary tool of the theory. It is not observable and as such can be eliminated from a description made only in terms of observable quantities.

We limit our considerations here to just one single example of spin-2 field theory that can be summarized in the following statements:
\begin{itemize}
	\item{Gravity is described by a symmetric second order tensor
		$\varphi_{\mu\nu}$ that satisfies a non-linear equation of motion;}
	\item{Matter couples to gravity in an universal way. In this interaction,
		the gravitational field appears only in the combination
		$g_{\mu\nu} = \gamma_{\mu\nu} + \varphi_{\mu\nu}$ where $\gamma_{\mu\nu}$ is the unobservable metric of the background. In general, it is associated to flat Minkowski structure. Such tensor
		$g_{\mu\nu}$ acts as the true metric tensor of the spacetime as seen by matter
		or energy of any form.}
\end{itemize}

In order to exhibit the complete covariance of the theory
all quantities will be described in an arbitrary system of
coordinates. In the auxiliary background geometry
of Minkowski spacetime of metric $\gamma_{\mu\nu}$ the
covariant derivative, represented by a semi-comma.
We define a three-index tensor $F_{\alpha\beta\mu}$, which we call the gravitational field, in terms of the symmetric standard variable
$\varphi_{\mu\nu}$ treated as the potential to describe
spin-two fields, by the expression
\begin{equation}
	F_{\alpha\mu\nu} = \frac{1}{2} ( \varphi_{\,\nu[\alpha;\mu]} +
	F_{\,[\alpha}\,\gamma_{\mu ]\nu} ),
	\label{d1}
\end{equation}
where $F_{\alpha}=$
In the expression above, the quantity $F_{\alpha}$ is the trace of $F_{\alpha\beta\mu}=F_{\alpha\mu\nu} \gamma^{\mu\nu}$, and the indices between square brackets indicates an antisymmetrization operation.
From the field variables we can form the following invariants
$A \equiv F_{\alpha\mu\nu}\hspace{0.5mm} F^{\alpha\mu\nu}~ \mbox{and} ~B \equiv F_{\mu}\hspace{0.5mm} F^{\mu}.$\footnote{Note
	that, besides these invariants, it is possible to define a quantity $C$, constructed with the dual, that is $C \equiv F^{*}_{\alpha\mu\nu}\, F^{\alpha\mu\nu}.$ We will not deal with such quantity here.}

\subsection{General relativity: the field formulation}
\label{ShortGR}

General relativity takes for granted that gravity is nothing but the fact
that all existing form of energy/matter
interacts through the modification of the universal geometry. However,
such a view is not exclusive and it is conceivable to try
to use two metrics to describe in an equivalent
way all content of such theory. There is no simpler and more direct way to
prove this statement than the one set forth by Feynmann. It is worth
to remark that a such duplication
causes no further difficulties when one realizes that the second
auxiliary metric $\gamma_{\mu\nu}$ is unobservable.

Let us pause for a while and make, just for completeness, a summary of the
principal features of this equivalent scheme.
The theory starts with the Fierz-Pauli linear equation
\begin{equation}
	G^{\mbox{\tiny (L)}}_{\mu\nu} = - \, \kappa\,T_{\mu\nu}
	\label{111}
\end{equation}
in which $T_{\mu\nu}$ is the matter energy-momentum tensor and
$G^{\mbox{\tiny (L)}}_{\mu\nu}$ is a linear tensor with respect to $\varphi_{\mu\nu}$ given by:
\begin{equation}
	G^{\mbox{\tiny (L)}}_{\mu\nu}\equiv \Box\varphi_{\mu\nu} -
	\varphi_{\ \mu ;\alpha\nu}^{\alpha}
	-\varphi_{\ \nu ;\alpha\mu}^{\alpha} +\varphi_{\ \alpha ;\mu\nu}^{\alpha}
	-\gamma_{\mu\nu}(\Box\varphi_{\ \alpha}^{\alpha}
	-\varphi^{\alpha\beta}_{\ \ ;\alpha\beta}).
\end{equation}
The action for this linear theory is $ S^{\mbox{\tiny (L)}} =  \int {\rm d}^{4}x\,
\sqrt{-\gamma}(A-B)$.
Since $G^{\mbox{\tiny (L)}}_{\mu\nu}$ is divergence-free it follows for
coherence that
the matter energy momentum tensor should also be divergence-free. However
this is in contradiction with the fact that gravity may exchange
energy with matter. To overcome such situation, one introduces an object
which we call Gupta-Feynman gravitational energy tensor $t^{(g)}_{\mu\nu} $
--- a cumbersome non linear expression in terms of $\varphi_{\alpha\beta}$
and its derivatives --- that is to be added to the right hand side of
Eq. (\ref{111}) in order to obtain a compatible set of equations,
\begin{equation}
	G^{\mbox{\tiny (L)}}_{\mu\nu} = - \, \kappa\,\left[ t^{(g)}_{\mu\nu}
	+ T_{\mu\nu}\right].
	\label{112}
\end{equation}

Note that, instead of using the standard procedure (as it happens in others
nonlinear theories) --- which in the case we examine here, asks for
the introduction of a nonlinear functional of the invariants $ A $ and $ B $,
dealt with in the linear case --- in order to obtain the dynamics of GR,
one must use other functionals of the basic field
$\varphi_{\mu\nu}$ which are not present in the linear case,
that means, they are not displayed in terms of the invariants $A$ and $B$.
We do not intend to repeat here the whole procedure (see Refs.
\cite{Deser,GPP} for more details),
but only to call the reader\rq s attention to such an unusual treatment of
dealing with a nonlinear process.
The origin of this approach goes back to the hypothesis
of the validity of the equivalence principle for gravitational energy.

For a convenient choice of the expression of the Gupta-Feynman gravitational energy, the equation (\ref{112}) is nothing but Einstein dynamics (see, for instance, Ref. \cite{GPP}). However, another path to deal with a nonlinear extension of Fierz original model is possible,  arriving at a different dynamics than in GR.

\subsection{A non-linear Fierz extension field theory of gravity}
\label{ShortNDL}

The non-linear Fierz extension (NLFE) starts at the same point as general relativity, that is, Fierz linear theory for spin-2 field. However, instead of breaking the symmetry displayed in the linear regime, presented in the combination of invariants under the form $ A - B $, as it was done in the Einstein case,  it assumes that this symmetry is maintained even after the introduction of non-linearities \cite{NDL}. The theory incorporates a great part of general relativity, it satisfies all standard tests of gravity and can be interpreted in the
standard geometrical way like GR, as far as the interaction of matter
to gravity is concerned. The most important particularity of the new theory concerns the gravity self interaction.

It is considered a more general Lagrangian function $L(U)$, with $U=A-B$.
Taking the variation of the gravitational action with
respect to the potential $\varphi_{\mu\nu}$, it results in the
following equations of motion,
\begin{equation}
	\left[L_{U} F^{\lambda (\mu\nu)} \right]_{;\lambda}
	= - \, T^{\mu\nu}
	\label{eqmov}
\end{equation}
where $L_{U}$ represents the derivative of the Lagrangian
with respect to the invariant $U$ and $T^{\mu\nu}$ is the energy-momentum tensor density of the matter contents.

A short analysis of the wave propagation description in this
theory shows that it satisfies,
\begin{equation}
	k^{\mu} k^{\nu} [ \gamma_{\mu\nu} + \Lambda_{\mu\nu} ] = 0,
	\label{51}
\end{equation}
where $k_{\alpha}$ represents the wave vector and
\begin{equation}
	\Lambda_{\mu\nu} \equiv 2 \frac{L_{UU}}{L_{U}}
	[ {F_{\mu}}^{\alpha\beta}
	F_{\nu (\alpha\beta)} - F_{\mu} F_{\nu}  ].
\end{equation}
Therefore, the gravitational disturbances propagate in a
modified geometry,
changing the background geometry $\gamma_{\mu\nu}$, into an effective
one $g_{\mu\nu}$, which depends on the energy distribution of the field
$F_{\alpha\beta\mu}$. This fact shows that such a property stems from
the structural form of the Lagrangian. Once the velocity of propagation of gravitational waves has been constrained to be the same as the light \cite{Abbott_2017}, this fact should strongly restrict the class of
Lagrangian functions $L(U)$.

\section{Final comments}\label{sec:conclusion}

The appearance of the method of the effective metric introduced by Gordon in 1923 and developed in the present century shows that the use of a Riemannian geometry, by general relativity, to describe gravity is one of many possibilities. The work of Feynmann, Grishchuk and others that used a spin-2 formulation in an unobserved space-time background has shown explicitly the similitude of the field formulation and the original metric description of general relativity.

Such field theory formulation of gravity is not restricted to a spin-2 formulation but instead can be described in terms of other structures, like a scalar or a spinor field. We have reviewed how recent advances in this direction has paved the way for new theoretical developments within the alternative gravity program, evidencing similarities and distinctions between those formulations and the paradigmatic general relativity. There is the possibility to go beyond Einstein-Nordstr\"{o}n's early scalar formulations of gravity. The single scalar field model shows a satisfactory description of local physics, gravitational radiation and cosmological scenarios where non-singular universes are natural solutions. A similar proposal using a non-linear spinor field satisfying Heisenberg dynamics uses the effective metric method to arrive at a gravitational theory in terms of fundamental fermion fields. This opens a new way to an unification between two different kinds of interactions, that is, gravity and the Fermi (weak) processes.

The method of effective metric and the consequent gravitational theories discussed here should also be viewed as a general framework for field theories of gravity. The theoretical models reviewed are specific cases of study within a rather much more enlarged scenario, and more general formulations still remains to be explored. In this sense, we hope that this work and the general picture here depicted can put on perspective several aspects of field formulations of alternative theories of gravity, motivating the future researches in the area.

\section*{Acknowledgements} MN would like to thank Dr. A. Hartmann for his interest and comments in a previous draft of this work, and Funda\c{c}\~{a}o de Amparo \`{a} Pesquisa do Estado do Rio de Janeiro (FAPERJ) for a fellowship.

\bibliography{bibToniato}

\end{document}